\documentclass[%
reprint,
superscriptaddress,
 amsmath,amssymb,
 aps,
]{revtex4-2}
\usepackage{xcolor}
\usepackage{graphicx}
\usepackage{dcolumn}
\usepackage{bm}

\usepackage{booktabs}

\begin{document}


\title{\boldmath Muonium Spectroscopy as a Quantum Sensor for Axion Dark Matter
}

\author{Feng Fang}
\affiliation{Advanced Energy Science and Technology Guangdong Laboratory, Huizhou 516000, China}	
\affiliation{Institute of Modern Physics, CAS, Lanzhou 730000, China}

\author{Kim Siang Khaw}
\affiliation{Key Laboratory for Particle Astrophysics and Cosmology (MOE), Shanghai Key Laboratory for Particle Physics and Cosmology (SKLPPC), Tsung-Dao Lee Institute \& School of Physics and Astronomy, Shanghai Jiao Tong University, Shanghai 201210, China}

\author{Ce Zhang}
\email{ce.zhang@liverpool.ac.uk}
\affiliation{University of Liverpool, Liverpool, United Kingdom}

\author{Qiaoli Yang}
\email{qiaoliyang@jnu.edu.cn}
\affiliation{Department of Physics, Jinan University, Guangzhou 510032, China}	

\author{Liangwen Chen}
\email{chenlw@impcas.ac.cn}
\affiliation{Institute of Modern Physics, CAS, Lanzhou 730000, China}
\affiliation{Advanced Energy Science and Technology Guangdong Laboratory, Huizhou 516000, China}	
\affiliation{State Key Laboratory of Heavy Ion Science and Technology, Institute of Modern Physics, Chinese
Academy of Sciences, Lanzhou 730000, China}
\affiliation{School of Nuclear Science and Technology, University of Chinese Academy of Sciences, Beijing 100049, China}

\author{Jie Yang}
\affiliation{Institute of Modern Physics, CAS, Lanzhou 730000, China}
\affiliation{Advanced Energy Science and Technology Guangdong Laboratory, Huizhou 516000, China}
\affiliation{State Key Laboratory of Heavy Ion Science and Technology, Institute of Modern Physics, Chinese
Academy of Sciences, Lanzhou 730000, China}
\affiliation{School of Nuclear Science and Technology, University of Chinese Academy of Sciences, Beijing 100049, China}

\author{Lei Yang}
\affiliation{Institute of Modern Physics, CAS, Lanzhou 730000, China}
\affiliation{Advanced Energy Science and Technology Guangdong Laboratory, Huizhou 516000, China}
\affiliation{State Key Laboratory of Heavy Ion Science and Technology, Institute of Modern Physics, Chinese
Academy of Sciences, Lanzhou 730000, China}
\affiliation{School of Nuclear Science and Technology, University of Chinese Academy of Sciences, Beijing 100049, China}

\author{Zhiyu Sun}
\email{sunzhy@impcas.ac.cn}
\affiliation{Institute of Modern Physics, CAS, Lanzhou 730000, China}
\affiliation{Advanced Energy Science and Technology Guangdong Laboratory, Huizhou 516000, China}
\affiliation{State Key Laboratory of Heavy Ion Science and Technology, Institute of Modern Physics, Chinese
Academy of Sciences, Lanzhou 730000, China}
\affiliation{School of Nuclear Science and Technology, University of Chinese Academy of Sciences, Beijing 100049, China}

\date{\today}

\begin{abstract}
High-intensity muon beams could enable a Muonium-based Axion Search through resonant quantum transitions between Hyperfine states (MASH). Combining theoretical calculations with simulation results, we demonstrate that such a muonium-based experimental approach could complement and tighten constraints on the axion-muon coupling beyond existing limits from the muon $g\!-\!2$ measurement, over the axion mass range of 18--130 $\mu$eV.
These results establish a new spectroscopic channel in muonium that enables searches for axion and axion-like particle dark matter.

\end{abstract}

\keywords{}

\maketitle


\textit{Introduction}—Astrophysical and cosmological observations indicate that approximately 27\% of the universe's total energy density consists of dark matter \cite{Planck:2018vyg}. Although the fundamental nature of dark matter remains poorly understood, both observational evidence and theoretical considerations suggest that dark matter cannot be composed of Standard Model (SM) particles. It is widely accepted that dark matter is non-relativistic and interacts weakly with SM particles \cite{Bond:1984fp}. Numerous theoretical models and dedicated experimental searches are currently underway to unravel this fundamental mystery.

Among the leading candidates for dark matter are weakly interacting massive particles (WIMPs) and axion/axion-like particles (ALPs). The coupling strength between ALPs and SM particles is primarily determined by the ALP decay constant. These particles are hypothesized to have been produced in the early universe and may constitute a significant fraction of dark matter. The extremely weak couplings of ALPs make their direct detection exceptionally challenging \cite{Marsh:2015xka,Sikivie:2020zpn,Gelmini:2016emn}. Nevertheless, any successful observation would represent a major breakthrough with profound implications.

As a purely leptonic system, muonium—a hydrogen-like bound state consisting of a positive muon and an electron—provides an ideal platform for precision tests of physics beyond the SM~\cite{Blumer:2024fvc,Zhang:2021cba,Tanaka:2021jtf,Stadnik:2022gaa}. Unlike hadronic systems, its simple two-body structure avoids complications from strong interaction effects, enabling highly precise theoretical predictions and more straightforward interpretation of experimental results. Consequently, the exceptional precision achievable in muonium spectroscopy offers significant potential for probing subtle new physical phenomena \cite{Delaunay:2021uph,Aiba:2021bxe}. Furthermore, the binary nature of muonium spin states provides a distinct advantage: rapid spin switching enables what amounts to a quasi-digital measurement scheme that can dramatically suppress systematic backgrounds through clear state discrimination. 

Muonium can be produced by injecting positive muons into gaseous targets \cite{PhysRevLett.82.711} or solid materials \cite{PhysRevLett.84.1136,PhysRevLett.128.011802}. High-intensity surface muon beams required for such production are typically generated at high-energy proton accelerator facilities. Surface muons are positive muons originating from pions decaying at rest on the target and have a kinetic energy of approximately 4 MeV. Several high-intensity muon sources currently operate worldwide, including ISIS, J-PARC, PSI, TRIUMF, and RCNP-MuSIC~\cite{Miyake:2010zz,Thomason:2019fwe,Maso:2023zjp,Marshall:1991wb,Stratakis:2017uci,Kanda:2020csq}. 
Presently, the most intense muon beam reaches intensities of approximately $10^{8} \mu ^{+} \mathrm{/s}$ at PSI. Several next-generation muon facilities are under development~\cite{Cai:2023caf,Liu:2025ejy,Bao:2023nup,KIM2020408,Xu:2025spd}, with projected capabilities that could elevate the muon intensity to $10^{10} \mu ^{+} \mathrm{/s}$ through the HIMB project \cite{Aiba:2021bxe}. Furthermore, the China Initiative Accelerator Driven System (CiADS), currently under construction, is expected to provide a surface muon beam with an unprecedented intensity of $5\times 10^{10} \mu ^{+} \mathrm{/s}$ \cite{Cai:2023caf}. The availability of such high-intensity muon beams will significantly expand opportunities for precision muonium experiments and enhance their sensitivity to potential new physics effects.

In this work, we present a comprehensive study of muonium as a precision probe for ALPs and propose an experimental design specifically optimized to measure ALP-induced transitions involving muons. While axion-electron and axion-photon couplings are already tightly constrained, the axion-muon coupling remains largely unexplored due to the muon's short lifetime and associated experimental challenges. The most stringent laboratory constraints currently derive from  the comparison between SM prediction and experimental measurements of the muon $g\!-\!2$ anomaly, with recent updates providing increasingly precise limits~\cite{Janish:2020knz, Athron:2025ets,Wu:2025jgk}. 
Muon $g\!-\!2$ measurements are sensitive to virtual axion or ALP loop effects, but the resulting bounds are model dependent, since additional couplings or new states can modify or cancel the loop contribution~\cite{Buen-Abad:2021fwq,Athron:2025ets}. 
Astrophysical limits, most notably from SN1987A, constrain axion emission from supernova cores and imply $g_{a\mu}<10^{-7.4}\,{\rm GeV}^{-1}$~\cite{PhysRevLett.125.051104,PhysRevLett.126.189901}; however, they rely on the modeling of dense nuclear matter, neutrino transport, and the muon abundance in the core~\cite{PhysRevD.101.123025}. Thus, muonium spectroscopy provides a complementary laboratory probe: it searches for resonant transitions driven by the real, oscillating axion dark-matter field in a controlled atomic system, independently of supernova modeling and of virtual-loop interpretations, opening a new laboratory window on a coupling sector poorly accessible to conventional searches.
\begin{figure} [ht]
    \centering
    \includegraphics[width=7 cm]{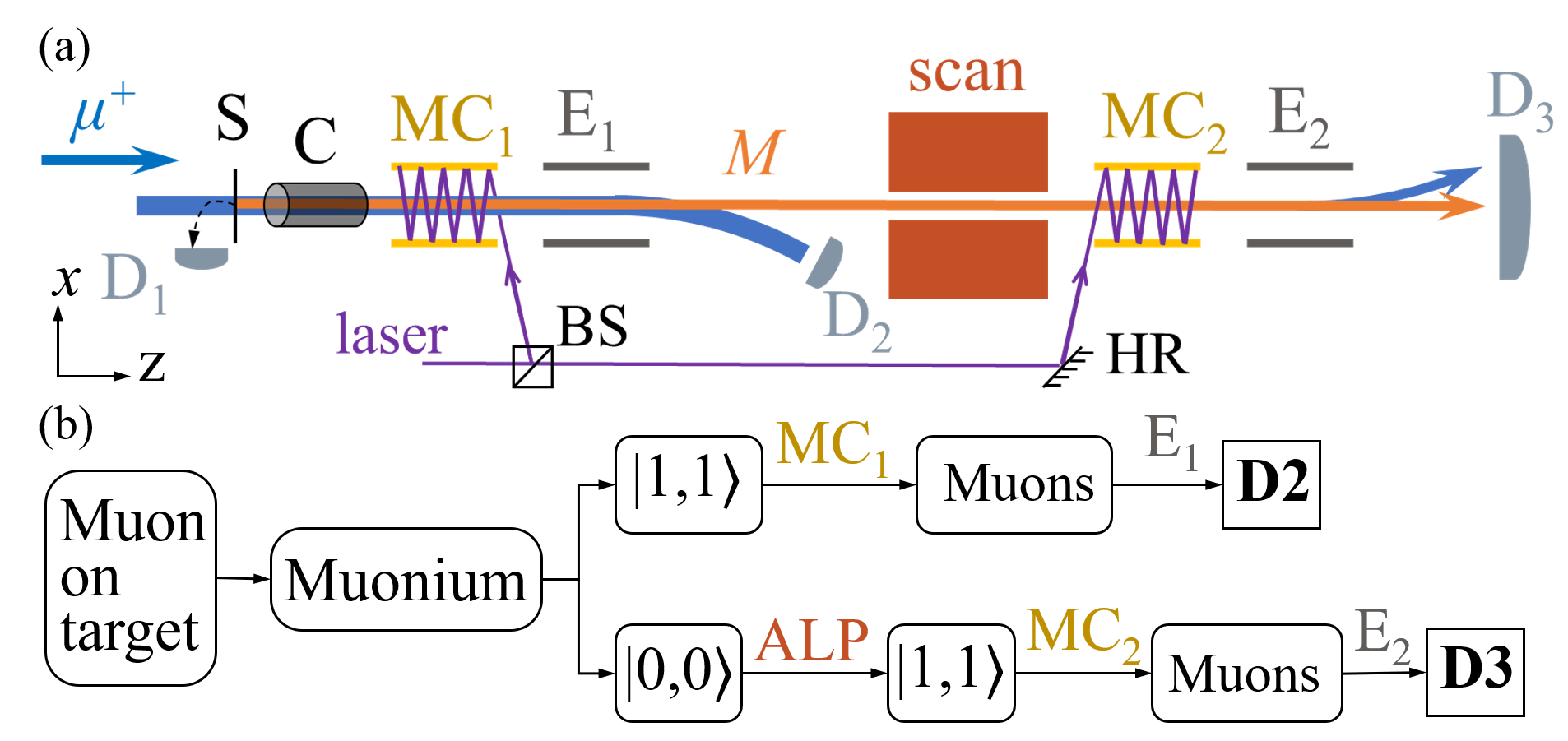}
    
    \caption{(a) Schematic diagram of the experimental setup. The muon spin direction is along the z direction. $\mathrm{S}$ and $\mathrm{C}$ represent a carbon foil and a collimator. $\mathrm{D_1}$, $\mathrm{D_2}$ and $\mathrm{D_3}$ are detectors. $\mathrm{E_1}$ and $\mathrm{E_2}$ are deflection electrodes with opposite polarities, which are designed to enhance the signal-to-background discrimination. The ``scan" region  denotes where the muonium atoms interact with ALPs and the magnetic field is oriented along the z direction. $\mathrm{MC_1}$ and $\mathrm{MC_2}$ represent multipass cavities. Laser beams consist of photons with wavelengths of 122 nm and 355 nm. BS and HR denote beam splitter and high reflection, respectively. (b) Diagram of experimental process.}
    
    \label{fig:geodrawing}
\end{figure}

\textit{Theoretical foundation}—The ALP field $a$ interacts with both muons and electrons through pseudoscalar couplings which are spin-dependent. By applying an external Zeeman magnetic field $B$, we can tune the splitting between hyperfine muonium states $|0,0\rangle$ and $|1,1\rangle$, creating an energy gap $\Delta E \approx 9.23+57.67B+9.23\sqrt{1+(6.31B)^2} ~\mu$eV, where $B$ is in tesla \cite{Tanaka:2021jtf,MuSEUM:2025cmo}. When this energy gap matches the ALP mass $m_a$, resonant quantum transitions occur at a rate proportional to the square of the axion-muon coupling strength. The induced signal events are \cite{Yang:2019xdz}
\begin{eqnarray}
(Rt_i)\times (\dot N T)\approx\pi g_{a\mu}^2L^2\rho_{DM}\dot N T,
\label{eq:NR}
\end{eqnarray}
where $R$ is the transition rate per muonium, $t_i$ is the interaction time for each muonium, $\dot N$ denotes the muonium intensity, and $T$ is the total measurement time. Additionally, $g_{a\mu}$ represents the ALP coupling to the muons, $L$ is the length of the magnetic field region, and $\rho_{\rm DM}$ is the local dark matter density. The spin flip of the muonium system provides a clear experimental signature that can be readily distinguished from backgrounds. 
Through selective ionization, the ALP-excited muonium atoms can be detected, resulting in an excess counting rate when the magnetic field is tuned to the ALP resonance condition.
A total of $\mathcal{O}(10)$ events from a potential signal would lead to a sensitivity to the ALP–muon coupling:
\begin{equation}
|g_{a\mu}| > {1\over L}\sqrt{ {\mathcal{O}(10)} \over \pi\rho_{DM}\dot N T }~~.
\label{eq:gau}
\end{equation}

Please see the End Matter part for technical details.

\textit{Experimental apparatus}—The experiment consists of three critical components: preparation of the muon beam, formation of muonium atoms, and detection of ALP-induced transitions, as shown in Fig.~\ref{fig:geodrawing}. 

The muon beam, which can be operated in either continuous or pulsed mode, is composed of surface muons. Upon passing through a cryogenic target, the muons are moderated to kinetic energies of a few keV before entering the experimental terminal. Then the moderated muon beam passes through a thin carbon foil, which emits several secondary electrons with energies of a few eV.
These electrons are extracted and guided by an electric field towards detector $D_1$, which provides a trigger for detector $D_3$. 
Detector $D_1$ could be composed of microchannel plates (MCPs), similar to the configuration adopted in Ref.~\cite{PhysRevLett.128.011802}.

Upon passing through the carbon foil, the muons capture electrons to form muonium atoms, which retain the forward kinetic energy of the incident muons. 
Over $90\%$ of the produced muonium atoms occupy the 1S state \cite{janka2020intense}, with the singlet and triplet hyperfine states statistically distributed \cite{PhysRevA.1.595}.

\begin{table}[htbp]
    \centering
    \caption{Energy splitting between the muonium states of $|0,0 \rangle$ and $|1,1 \rangle$ under different magnetic fields, calculated using the Breit-Rabi formula.}
    \begin{tabular}{lccc}  
        \toprule
        B (T) &  $0$ &  $0.1$ &  $1.0$ \\
        \midrule
        E (eV) & $1.8\times 10^{-5}$ & $2.6\times 10^{-5}$ &  $1.3\times 10^{-4}$  \\
        \bottomrule
    \end{tabular}
    \label{tab:T1}
\end{table}

The muonium atoms subsequently enter the first laser cavity, where atoms in the $|1,1 \rangle$ state are selectively removed, while those in the $|0,0 \rangle$ ground state remain unaffected. A $122\ \mathrm{nm}$ laser is tuned to resonantly excite the transition of $1S_{1/2},\, F=1,\, m_F=1$ to $2P_{3/2},\, F=2,\, m_F=2$. Meanwhile, a 355 nm laser ionizes the $2P_{3/2}$ atoms. The state population can be estimated using rate equations \cite{Nakajima:12} for an excitation duration of 2 ns. A 122 nm laser pulse with an energy of 100 $\mu$J \cite{Saito:16}, pulse width of 2 ns and beam waist of 1 cm can provide an excitation probability exceeding 95\%. To suppress unwanted excitation of the $|0,0 \rangle$ state, the linewidth of the 122 nm laser must remain below 4.5 GHz in the absence of a significant magnetic field. 
Immediately, muonium atoms enter the downstream electric field region $E_1$, where the neutral atoms pass straight through, while the bare muons are deflected into the detector $D_2$. It is also composed of MCPs for the detection of free muons \cite{KIM201822}. In this way, the incoming muonium atoms in the $|1, 1 \rangle$ state are selectively destroyed, whereas those in the $|0, 0 \rangle$ state are unaffected.



Subsequently, the surviving muonium atoms in the $|0, 0 \rangle$ state enter the scan region, which constitutes the core of the experiment. This region is composed of a solenoid magnet aligned along the z direction. The magnet configuration is carefully designed to ensure a fluctuation of magnetic field less than 0.3 G~\cite{KIM201822}, for matching the transit-time broadening of 0.5 MHz. Meanwhile, the magnetic field is maintained without crossing zero throughout the transport region in order to suppress non-adiabatic transitions between hyperfine substates.
With a magnetic field of around 1 T, the Zeeman splitting energy between the $|0,0 \rangle$ and $|1, 1 \rangle$ is tens of $\mu$eV as shown in Table \ref{tab:T1}.
A resonant transition to the $|1, 1 \rangle$ state would occur if this energy splitting matches the mass of the ALPs.

After that, the ALP-excited atoms are selectively ionized by the lasers in a second multipass cavity. The resulting muons are then guided to the detector $D_3$, which could serve as a definitive signature of ALPs. Detector $D_3$ could be implemented as a channel electron multiplier with nearly 100\% detection efficiency compared with MCPs.

\begin{figure}[ht]
\centering
\includegraphics[width=7 cm]{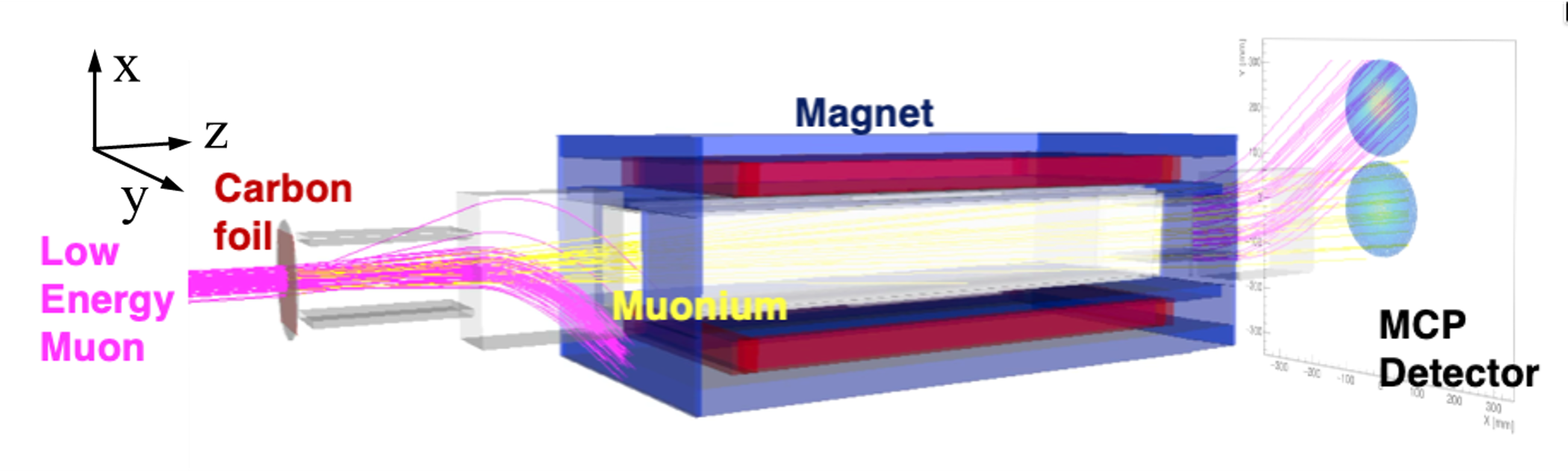}
\caption{Experimental model constructed in Geant4 with an illustrated signal profile at the MCP.}
\label{fig:setup}
\end{figure}

\textit{Sensitivity simulation}—To validate the experimental design and optimize key parameters, we performed detailed simulations using the musrSim package~\cite{sedlak_musrsim_2012}, which is based on Geant4~\cite{GEANT4:2002zbu}, as shown in Fig.~\ref{fig:setup}. The model includes a 10 nm thick carbon foil target, a few-meter-long scan magnets, laser cavities, deflection electrodes, and MCP detectors, in accordance with Fig.~\ref{fig:geodrawing}. Muonium formation was simulated through muon interactions in the carbon foil, taking into account energy loss and secondary electron emission. This procedure has been benchmarked against previous work at PSI~\cite{Khaw:2015eya, Antognini:2011ei}.

The performance of the setup was first quantified under representative conditions. The incident muon beam was defined with a transverse vertex width of $7.5\ \rm mm$ and a beam divergence of $\sigma_{x'/y'}\approx2.0^{\circ}$, consistent with Table 2 of Ref.\cite{Khaw:2015eya}. Multiple scattering in the carbon foil dominated the emittance of the outgoing muonium beam, which in turn constrained the achievable spot size at the MCP. A laser cavity of dimensions $20\times20\times1\ \rm cm^3$ along $x$, $y$, and $z$, respectively, was applied as a representive fiducial region for muonium tracks, sufficiently large to encompass the emittance growth toward the MCP. A moderate electric field of 0.03 kV/mm placed before the scan magnet provided sufficient deflection to separate the penetrating muon background from the undeflected muonium. Following the second laser cavity, a 0.1 kV/mm electric field was applied and the ALP-induced muons were consequently deflected by more than 20 cm toward the MCP detector.

\begin{figure}[ht]
\centering
\includegraphics[width=7.5 cm]{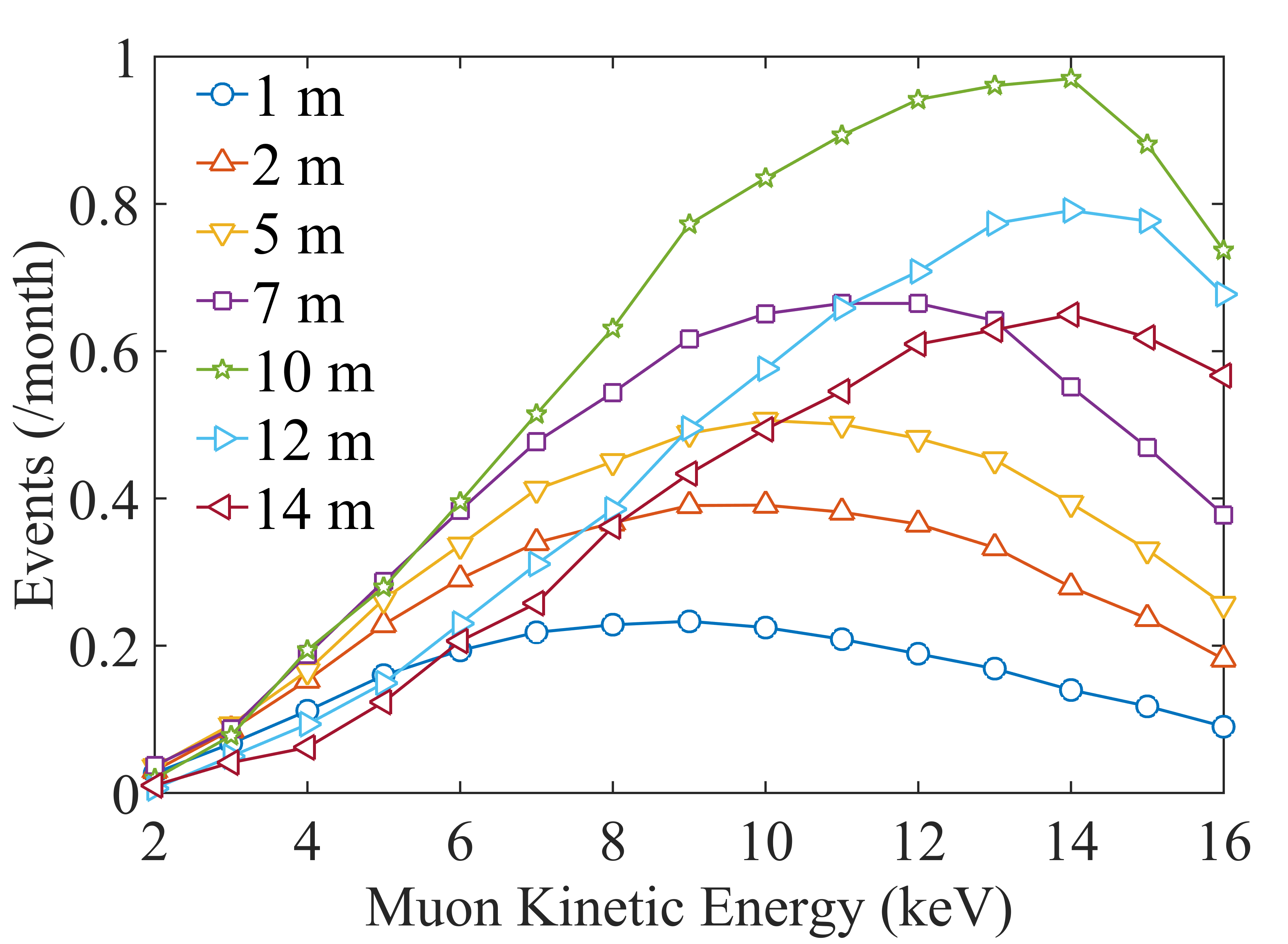}
\caption{Dependence of the detected ALP-induced events on the incident muon kinetic energy and magnetic field length $L$ spanning $1$--$14\ \mathrm{m}$, evaluated for $m_a = 2 \times 10^{-5}\ \mathrm{eV}$ with the corresponding sensitivity of $g_{a\mu} = 2.6\times 10^{-12}\ \mathrm{eV^{-1}}$. The signal arises from muonium atoms undergoing ALP-induced transitions followed by laser ionization.}
\label{fig:signal_rate}
\end{figure}

Assuming a realistic slow muon rate of $10^4$/s (from a surface muon rate of $10^8$/s), the yield muonium transported to the MCP bending direction is about $4.1\times10^2$/s, including the expected $\sim$10\% muonium formation probability at the carbon foil for $\sim$10 keV muons~\cite{Khaw:2015eya}, as well as decay and transition losses. For a future scenario with $5\times10^{10}$/s surface muons available at facilities such as PSI, the corresponding slow muon rate of $3\times10^6$/s would increase the muonium rate by two orders of magnitude.
We performed a two-parameter scan over the injection energy and the magnetic length $L$, assuming the above enhanced rate, a local dark matter density of $\rho_{\rm DM}=0.3\ \mathrm{GeV/cm^3}$, and a coupling strength of $g_{a\mu}=2.6\times10^{-12}\ \mathrm{eV}^{-1}$, as shown in Fig.~\ref{fig:signal_rate}. The optimization reveals a trade-off: lower energy enhances conversion but increases decay losses due to reduced velocity, while longer $L$ extends interaction time but also amplifies beam divergence. 
Using Eq.~\ref{eq:NR}, the resulting maximum occurs near $14~\mathrm{keV}$ with $L=10~\mathrm{m}$, where the predicted signal rate reaches $\sim 1$ per month for ALP masses of $\mathcal{O}(10^{-5})\ \mathrm{eV}$.






Several potential background components are considered and simulated, including prompt photons, penetrating muons, muonium atoms, and positrons from muon decays or scattering. Among these sources, photons and the very small number of penetrating muons can be efficiently rejected by applying time-of-flight (TOF) discrimination. Muonium atoms and positrons are further suppressed by the bending electric field together with the spatial selections described above, whose optimized values reduce both contributions to a negligible level. 
The dominant background originates from residual muonium atoms already in the $|1,1 \rangle$ state that survive the first laser-ionization stage due to the finite cleaning efficiency. The sensitivity is primarily limited by the stability of the residual background. A resonance-modulation strategy is necessary based on the narrow bandwidth of the ALP-induced hyperfine transition. Because the signal appears only under the resonance condition while the leakage background remains independent of the magnetic-field tuning, a differential on/off-resonance analysis can strongly suppress the effective background contribution. The narrow resonance bandwidth of $\Delta\nu/\nu\sim 10^{-5}$ reduces the relevant background within the signal window by the same factor, allowing an expected signal yield of $\mathcal{O}(10)$ events/year to reach observable significance. Owing to the low effective muonium density in the scan region, the  collision probability and collective stray-field effects from decay particles are expected to be negligible.





\begin{figure}
    \centering
    \includegraphics[width=7 cm]{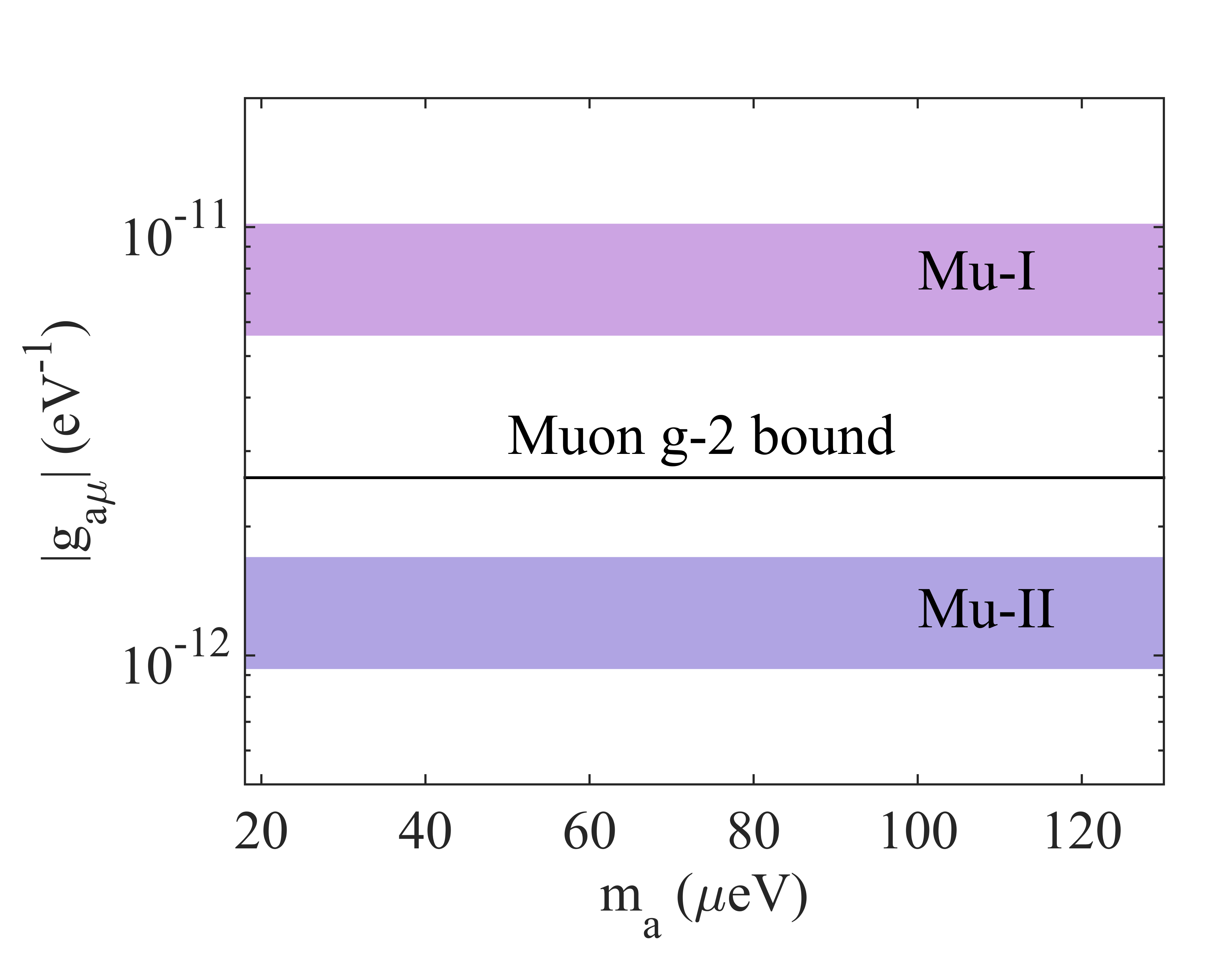}
    \caption{Projected sensitivity of the proposed scheme with one-month (Mu-I) and three-year (Mu-II) data. The upper and lower limits of each region correspond to the local dark matter densities $\rho_{\text{DM}}$ of $0.3\ \text{GeV/cm}^3$ and $1.0\ \text{GeV/cm}^3$, respectively. The solid black line shows the constraints from muon $g\!-\!2$ bound \cite{Buen-Abad:2021fwq,Athron:2025ets,Aliberti:2025beg,Muong-2:2025xyk}.}
    \label{fig:constraints}
\end{figure}    

Figure \ref{fig:constraints} shows the projected sensitivity calculated according to Eq.~\ref{eq:gau}, assuming $L = 10~\mathrm{m}$ and $\dot N=2\times 10^5~\mathrm{s^{-1}}$. For one month of data acquisition, the expected sensitivity approaches the current limit deduced from the discrepancy between the experimental measurement~\cite{Muong-2:2025xyk} and theoretical prediction of the muon $g\!-\!2$~\cite{Aliberti:2025beg}. Although the muon $g\!-\!2$ bound provides the strongest existing constraints obtained in terrestrial experiments, it is indirect and therefore subject to model-dependent interpretations. By accumulating larger statistics, the proposed direct ALP search experiment has the potential to surpass the current muon $g-2$ bound in the relevant parameter region.

\textit{Conclusions}—We have proposed and simulated a novel muonium-based experiment to search for axion-like particles through quantum transitions. Using the Geant4 simulation framework, we validated the feasibility of the design, demonstrated effective background suppression, and identified optimal beam and field parameters. 
The combination of dual electric-field deflection, laser ionization, and selective MCP detection enables an essentially background-free measurement. With the advent of next-generation high-intensity muon sources (such as HiMB, CiADS), our projected sensitivity can surpass the most stringent terrestrial constraints for $m_a$ in the range of 18--130 $\mu$eV. This work thus opens a new window on the ALP–muon couplings---a sector that has been largely inaccessible to colliders and precision muon experiments due to the pseudoscalar nature of the interaction.

\textit{Acknowledgements}—This work is financially supported by the National Natural Science Foundation of China (Grant No. 12105327), the Guangdong Basic and Applied Basic
542 Research Foundation (Grant No. 2023B1515120067), the Research Program of State Key Laboratory of Heavy Ion Science and Technology, Institute of Modern Physics, Chinese Academy of Sciences, Lanzhou 730000, China, under Grant No. HIST2025CS06. This work is also supported by High Intensity heavy-ion Accelerator Facility (HIAF) project approved by the National Development and Reform Commission of China. Computational resources were provided by the Dongjiang Yuan Intelligent Computing Center supercomputing facility. The work of C. Z. was supported by the Leverhulme Trust, LIP-2021-014.

\textit{Data availability}—The simulation data used to produce the results presented in this Letter are not publicly available upon publication because it is not technically feasible and/or the cost of preparing, depositing, and hosting the data would be prohibitive within the terms of this research project. The data are available from the authors upon reasonable request.

\bibliography{ref.bib}

\onecolumngrid
\section*{E\lowercase{nd} M\lowercase{atter}}
\twocolumngrid

 ALPs interact with the muon and electron via:  
\begin{equation}
\mathcal{L}_{\text{int}} = -\sum_{l=\mu,e} \frac{g_l}{f_a} \partial_\mu a \, \bar{\psi}_l \gamma^\mu \gamma^5 \psi_l,
\end{equation}
where \(g_l\) are model-dependent couplings of order one, \(f_a\) is the axion decay constant and the axion-lepton coupling defined as $g_{al}=g_l/f_a$, \(\psi_l\) represent the fermion fields, and $a$ denotes the ALP field. For non-relativistic muons and electrons, as in muonium, these conditions lead to the effective ALP-lepton interaction \cite{Stadnik:2013raa}:  
\begin{equation}
H_{\text{int}} \approx \sum_{l=\mu,e}  g_{al}m_a  \vec{\sigma}_l \cdot \vec{v}_a \, a_0 \sin(\omega_a t),
\end{equation}
where \(\vec{v}_a \) represents the ALP velocity relative to stationary leptons, and is of order $10^{-2}$c in this experimental setup. \(a_0 = \sqrt{2\rho_{\text{DM}}}/m_a\) denotes the ALP field amplitude, \(\rho_{\text{DM}} \approx 0.3\sim 1.0 \, \text{GeV/cm}^3\) is the local dark matter density \cite{PhysRevD.33.889, Sikivie:2020zpn}, and \(\omega_a \approx m_a (1 + v_a^2/2)\) is the center angular frequency of the dark matter field.

When the Zeeman splitting energy gap $\Delta E \approx 9.23+57.67B+9.23\sqrt{1+(6.31B)^2} ~\mu$eV
matches the ALP mass \(m_a\), resonant transitions occur between the states \(m_F=0\) and \(m_F=1\). The corresponding transition rate is given by:  
\begin{equation}
R = \pi \left| \sum_{l=\mu,e} g_{al} \langle f | \vec{\sigma}_l \cdot \vec{v}_a | i \rangle \right|^2 I_a,
\end{equation}  
where 
\begin{equation}
I_a = {\rho_{\text{DM}} \over {\rm Max}(m_a v_a\delta v,~1/t_i,~1/\tau_\mu)}
\end{equation}
represents the effective ALP spectral density during the interaction, $t_i$ is the interaction duration, $\tau_\mu$ is the lifetime of muonium and \(\delta v \sim 10^{-3}c\) is the dark matter halo velocity dispersion. The nonzero matrix element of a muon-philic axion can be written as 
\begin{equation}
\left| \sum_{l=\mu,e} g_{al} \langle f | \vec{\sigma}_l \cdot \vec{v}_a | i \rangle \right|^2\approx g_{a\mu}^2v_a^2~.
\end{equation}
In this specific experiment, we have $(m_av_a\delta v)^{-1}>10^{-5}$s, $t_i\approx2.0\times 10^{-6}$s, $\tau_\mu\approx2.2\times 10^{-6}$s, thus $I_a=\rho_{\rm DM}t_i$. This leads to the transition rate:
\begin{equation}
R\approx \pi g_{a\mu}^2v_a^2\rho_{\rm DM}t_i~.
\end{equation}
The event number per muonium atom is then $Rt_i$. Denote $L$ as the length of the
magnetic field in the interaction region. Because the muonium beam speed ($10^{-2}$c) is much larger than the axion halo velocity ($10^{-3}$c), the axion wind speed $|v_{a}|\approx|v_{\mu}|$. For a measurement running for a time $T$ and a muonium rate $\dot N$, the total number of muonium atoms passed the Zeeman field is $\dot NT$. Thus the total events during one measurement is 
\begin{eqnarray}
Rt_i\times \dot N T=\pi g_{a\mu}^2L^2\rho_{DM}\dot NT.
\end{eqnarray}  
The optimal scanning frequency bandwidth per bin is $\Delta f\approx {1\over t_i}\sim 0.5\times 10^6$ Hz. Considering a total of 3 events for a potential discovery, the projected sensitivity of $g_{a\mu}$ satisfies: 
\begin{equation}
|g_{a\mu}| >{1\over L} \sqrt{3\over \pi\rho_{DM}\dot N T} \approx9.6\times 10^{-12} {\rm eV^{-1}}~,
\end{equation}  
where $T=1$ month and $\dot N=2\times 10^{5}$/s.

\end{document}